\documentstyle[preprint,aps]{revtex}

\begin{document}
\title{Transverse Ward-Takahashi Relation for the Fermion-Boson
Vertex Function in 4-dimensional QED}

\author{Han-xin He $^{a,b}$ $^{*}$ \footnotetext{$^*$ E-mail address: hxhe@iris.ciae.ac.cn}}
\address{$^{a}$ China Institute of Atomic Energy, P.O.Box 275(18), Beijing
102413, P.R.China \\
$^{b}$ Institute of Theoretical Physics, the Chinese Academy of Science, Beijing 100080,
 P.R.China }

\maketitle
\begin{abstract}

I present a general expression of the transverse Ward-Takahashi
relation for the fermion-boson vertex function in momentum space
in 4-dimensional QED, from which the corresponding one-loop
expression is derived straightforwardly. Then I deduce carefully
 this transverse Ward-Takahashi relation to one-loop order in
d-dimensions, with $d = 4 + \epsilon$. The result shows that this
relation in d-dimensions has the same form as one given in
4-dimensions and there is no need for an additional piece
proportional to $(d-4)$ to include for this relation to hold in
4-dimensions. This result is confirmed by an explicit computation
of terms in this transverse WT relation to one-loop order. I also
make some comments on the paper given by Pennington and Williams
who checked the transverse Ward-Takahashi relation at one loop
order in d-dimensions.

\noindent PACS number(s): 11.30.-j; 11.15.Tk; 12.20.Ds.


\end{abstract}

\newpage

\section{Introduction}
The gauge symmetry imposes powerful constrains on the basic
interaction vertices, leading to a variety of exact relations
among Green's functions--- referred to as Ward-Takahashi(WT)
relations[1]. They play an important role in the study of gauge
theories through of the use of Dyson-Schwinger equations[2-5]. The
well-known and simplest WT relation relates the
fermion-gauge-boson vertex $\Gamma _V^\mu$ to the fermion
propagator $S_F$ :
\begin{equation}
q_\mu \Gamma _V^\mu (p_1,p_2)=S_F^{-1}(p_1)-S_F^{-1}(p_2),
\end{equation}
where $q=p_1-p_2$. The normal WT identity (1) is satisfied both
perturbatively and nonperturbatively, but it specifies only the
longitudinal part of the vertex, leaving the transverse part
unconstrained. It has long been known that the transverse part of
the vertex plays the crucial role in ensuring multiplicative
renormalizability and so in determining the propagator [2-5]. How
to determine the transverse part of the vertex then becomes a
crucial problem. Although much effort has been devoted to
constructing the transverse part of the vertex in terms of an
$ansatz$ which satisfies some constraints but all such attempts
remain $ad$
 $hoc$[2-6]. Such a constructed vertex is not unique since it is not
fixed by the symmetry of the system. Takahashi first discussed the
constraint relation for the transverse part of the vertex from
symmetry, which is called the transverse WT relation[7]. So
 far a basic formula of the transverse WT relation for the
fermion-boson vertex in coordinate space has been given as[8]

\begin{eqnarray}
& &\partial _x^\mu \left\langle 0\left| Tj^\nu (x)\psi
(x_1)\bar{\psi}(x_2) \right| 0\right\rangle -\partial _x^\nu
\left\langle 0\left|
Tj^\mu (x)\psi (x_1)\bar{\psi}(x_2)\right| 0\right\rangle \nonumber \\
&=&i\sigma ^{\mu \nu }\left\langle 0\left| T\psi
(x_1)\bar{\psi}(x_2) \right| 0\right\rangle \delta
^4(x_1-x)+i\left\langle 0\left| T\psi (x_1)
\bar{\psi}(x_2)\right| 0\right\rangle \sigma ^{\mu \nu }\delta ^4(x_2-x)\nonumber \\
& &+2m\left\langle 0\left| T\bar{\psi}(x)\sigma ^{\mu \nu }\psi
(x)\psi (x_1)
\bar{\psi}(x_2)\right| 0\right\rangle \nonumber \\
& &+{\lim _{x^{\prime }\rightarrow x}}i(\partial _\lambda ^x-
\partial _\lambda ^{x^{\prime }})\varepsilon ^{\lambda \mu \nu \rho }
\left\langle 0\left| T\bar{\psi}(x^{\prime })\gamma _\rho \gamma
_5 U_P (x^{\prime },x)\psi (x)\psi (x_1)\bar{\psi}(x_2)\right|
0\right\rangle ,
\end{eqnarray}
where $j^{\mu}(x)=\bar{\psi}(x)\gamma ^{\mu }\psi (x)$,
 $\sigma ^{\mu \nu }=\frac{i}{2}[\gamma ^{\mu },\gamma ^{\nu }]$ and
 $m$ is the bare fermion mass.
The Wilson line $U_P (x^{\prime },x)=P\exp (-ig\int_x^{x^{\prime
}}dy^\rho A_\rho (y))$ is introduced in order that the operator is
locally gauge invariant, where $A_{\rho} $ are gauge fields and $g
= e$ in QED case. The last matrix element in Eq.(2) expresses a
non-local axial-vector vertex function in coordinate space.

The momentum space representation of this transverse WT relation
is obtained by computing the Fourier transformation of Eq.(2),
which gives[10]:
\begin{eqnarray}
& &iq^\mu \Gamma _V^\nu (p_1,p_2)-iq^\nu \Gamma _V^\mu (p_1,p_2)\nonumber \\
&=&S_F^{-1}(p_1)\sigma ^{\mu \nu }+\sigma ^{\mu \nu }S_F^{-1}(p_2)+
2m\Gamma _T^{\mu \nu }(p_1,p_2)\nonumber \\
& &+(p_{1\lambda }+p_{2\lambda })\varepsilon ^{\lambda \mu \nu \rho }
\Gamma _{A\rho }(p_1,p_2)
-\int \frac{d^{4}k}{(2 \pi)^{4}}2k_{\lambda}
\varepsilon ^{\lambda \mu \nu \rho }\Gamma _{A\rho }(p_1,p_2;k),
\end{eqnarray}
where $\Gamma _{A\rho }$ and $\Gamma _T^{\mu \nu }$ denote the
axial-vector and tensor vertex functions, respectively, and the
integral-term involves the non-local axial-vector vertex function
$\Gamma _{A\rho}(p_1,p_2;k)$, with the internal momentum $k$ of
the gauge boson appearing in the Wilson line. This integral-term
was missing in the earlier works [8,9]. In perturbation theory the
integral-term at one-loop order is easy to write

\begin{eqnarray}
& &\int \frac{d^{4}k}{(2 \pi)^{4}}2k_{\lambda}
\varepsilon ^{\lambda \mu \nu \rho }\Gamma _{A\rho }(p_1,p_2;k)\nonumber \\
&=& g^2\int \frac{d^{4}k}{(2 \pi)^{4}}2k_{\lambda}
\varepsilon ^{\lambda \mu \nu \rho }\gamma^{\alpha}
\frac{1}{{\makebox[-0.8 mm][l]{/}{p}}_1-
{\makebox[-0.8 mm][l]{/}{k}}-m}\gamma_{\rho}\gamma_5
\frac{1}{{\makebox[-0.8 mm][l]{/}{p}}_2-{\makebox[-0.8 mm][l]{/}{k}}-m}\gamma^{\beta}
\frac{-i}{k^2}[g_{\alpha\beta}+(\xi-1)\frac{k_{\alpha}k_{\beta}}{k^2}]\nonumber \\
& &+g^2\int \frac{d^{4}k}{(2 \pi)^{4}}2
\varepsilon ^{\alpha \mu \nu \rho }[\gamma^{\beta}
\frac{1}{{\makebox[-0.8 mm][l]{/}{p}}_1-{\makebox[-0.8 mm][l]{/}{k}}-m}\gamma_{\rho}\gamma_5
+ \gamma_{\rho}\gamma_5
\frac{1}{{\makebox[-0.8 mm][l]{/}{p}}_2-{\makebox[-0.8 mm][l]{/}{k}}-m}\gamma^{\beta}]
\frac{-i}{k^2}[g_{\alpha\beta}+(\xi-1)\frac{k_{\alpha}k_{\beta}}{k^2}]
,
\end{eqnarray}
where ${\makebox[-0.8 mm][l]{/}{k}}=\gamma_{\mu}k^{\mu}$, $\xi$ is
the covariant gauge parameter. Here the integral-term involves two
parts: the first part is the one-loop axial-vector vertex
contribution, and the second part is the one-loop self-energy
contribution accompanying the vertex correction.

The transverse WT relation to one-loop order given by Eq.(3) with
Eq.(4) was derived in 4-dimensions[10] and was demonstrated to be
satisfied to one-loop order by Refs.[11,12] in the Feynman gauge
but without performing the check in d-dimensions, with $d=
4+\epsilon$. Especially, so far the complete expression of the
integral-term has not been given.

Recently, Pennington and Willams made a good comment on the
potential of the transverse WT relation to determine the full
fermion-boson vertex and then checked the transverse WT relation
to one loop order in d-dimensions[13]. However, they claimed that
an additional piece, say $(d-4)N^{\mu\nu}$, must be included in
the evaluation of the integral-term for this transverse WT
relation to hold in 4-dimensions. This problem is crucial and so
is worth to clarify for the further study and the application of
this transverse WT relation. Obviously, the central subject is
attributed to the study of the integral-term given in Eq.(3).

In this paper, at first, I present a complete expression of the
integral-term involving the non-local axial-vector vertex function
and hence give the general formula of the transverse WT relation
for the fermion-boson vertex in momentum space in 4-dimensional
QED, from which the corresponding one-loop expression can be
derived straightforwardly. To see if an additional piece
proportional to $(d-4)$ should be included in Eq.(3), I then
deduce carefully the transverse WT relation to one-loop order in
d-dimensions and compute explicitly the terms of this transverse
WT relation. The result shows that this transverse WT relation in
d-dimensions has the same form as one, Eq.(3) with Eq.(4), given
in 4-dimensions and there is no need for an additional piece $\sim
(d-4)$ to include for this relation to hold in 4-dimensions. The
complete expression of the integral-term and the detailed deducing
in d-dimensions are given in Sec.II. The one-loop result is
checked, by an explicit computation of terms in this relation, in
Sec.III. In Sec.IV, I show how the authors of Ref.[13] separate
out a so-called additional piece $(d-4)N^{\mu\nu}$ from the
integral-term given by Eq.(4) by defining a modifying
integral-term, which in fact does not change the original formula
of the transverse WT relation to one-loop order. The conclusion
and remark are given in Sec.V.

\section{Transverse Ward-Takahashi Relation in d-dimensions}

At first, let me write the complete expression of the
integral-term involving $\Gamma_{A\rho}(p_{1}, p_{2};k)$ in the
transverse WT relation (3), where $\Gamma_{A\rho}(p_{1}, p_{2};k)$
is given by the Fourier transformation of the last matrix element
in Eq.(2):
\begin{eqnarray}
& &\int d^{4}x d^{4}x^{\prime}d^{4}x_{1} d^{4}x_{2}
e^{i(p_{1}\cdot x_{1} - p_{2}\cdot  x_{2} + (p_2-k)\cdot x -
(p_1-k)\cdot x^{\prime})} \langle 0|T \bar{\psi}(x^{\prime
})\gamma _\rho \gamma _5 U_P (x^{\prime },x)\psi (x)\psi
(x_1)\bar{\psi}(x_2) |0\rangle \nonumber \\
&=& (2 \pi)^{4} \delta^{4}(p_{1} - p_{2} - q) iS_{F}(p_{1})
\Gamma_{A\rho}(p_{1}, p_{2};k) iS_{F}(p_{2}) ,
\end{eqnarray}
where $q=(p_1-k)-(p_2-k) $.  Eq.(3), together with Eq.(5), gives
the general expression of the transverse WT relation for the
fermion-boson vertex function in momentum space in 4-dimensional
QED, which should be satisfied both perturbatively and
non-perturbatively like the normal WT identity (1). In fact,
perturbative calculations of Eq.(5) can be performed order by
order in the interaction representation, which, at one-loop order,
leads straight to the expression given by Eq.(4).

Eq.(5) shows that $\Gamma_{A\rho}(p_{1}, p_{2};k)$ is a non-local
axial-vector vertex function, which and hence the integral-term is
the four-point-like function. This integral-term is essential for
this transverse WT relation to be satisfied both perturbatively
and non-perturbatively. Indeed, as shown by Refs.[11,12], this
integral-term is crucial to prove the transverse WT relation being
satisfied to one-loop order in perturbation theory.

In the following, let me check if the transverse WT relation to
one-loop order, given by Eq.(3) with Eq.(4), holds in 4-dimensions
by deducing  this transverse WT relation to one-loop order in
d-dimensions with a similar procedure outlined in Ref.[11].

The fermion-boson vertex to one-loop order in perturbation theory
is known as:
\begin{equation}
\Gamma _V^\mu(p_1,p_2) = \gamma^\mu + \Lambda^{\mu}_{(2)}(p_1,p_2)
\end{equation}
with
\begin{equation}
\Lambda^{\mu}_{(2)}(p_1,p_2) = \int \overline{d k}\gamma ^{\alpha}
{\makebox[-0.8 mm][l]{/}{a}}\gamma^{\mu} {\makebox[-0.8
mm][l]{/}{b}}\gamma_{\alpha},
\end{equation}
where the shorthand denotes
\begin{equation}
\int \overline{d k} = g^2\int \frac{d^{d}k}{(2 \pi)^{d}}
\frac{-i}{k^2a^2b^2 } ,\ \ \ a=p_1-k, b=p_2-k,
\end{equation}
and the case of massless fermion and the Feynman gauge are taken
for simplifying the discussion. Eq.(6) gives
\begin{equation}
iq^\mu \Gamma _V^\nu (p_1,p_2)-iq^\nu \Gamma _V^\mu (p_1,p_2) =
i(q^\mu \gamma ^\nu - q^\nu \gamma ^\mu) + i(q^{\mu}
\Lambda^{\nu}_{(2)}(p_1,p_2)- q^{\nu}
\Lambda^{\mu}_{(2)}(p_1,p_2)).
\end{equation}
Here $i(q^\mu \gamma ^\nu - q^\nu \gamma ^\mu)$ satisfies the
transverse WT relation (3) at tree level, which is reduced to a
trivial identity of $\gamma$-matrices:
\begin{equation}
iq^\mu \gamma ^\nu -iq^\nu \gamma ^\mu = {\makebox[-0.8
mm][l]{/}{p}}_1\sigma ^{\mu \nu }+\sigma ^{\mu \nu }
{\makebox[-0.8 mm][l]{/}{p}}_2 +(p_{1\lambda }+p_{2\lambda
})\varepsilon ^{\lambda \mu \nu \rho } \gamma _{\rho }\gamma_5 .
\end{equation}
Using Eq.(7)and Eq.(10) and the identity of $\gamma$-matrices,
\begin{equation}
ir^{\lambda \mu \nu} = \{\gamma ^{\lambda}, \sigma ^{\mu \nu}\} =
- 2\varepsilon ^{\lambda \mu \nu \rho } \gamma _{\rho}\gamma_5 = i
(\gamma ^{\lambda}\gamma^{\mu}\gamma^{\nu} -
\gamma^{\nu}\gamma^{\mu}\gamma ^{\lambda} ),
\end{equation}
the last term in the right-hand side
of Eq.(9) can be written as
\begin{eqnarray}
& &iq^{\mu} \Lambda^{\nu}_{(2)}(p_1,p_2)
- iq^{\nu} \Lambda^{\mu}_{(2)}(p_1,p_2) \nonumber \\
&=&\int \overline{d k} \gamma^{\alpha} {\makebox[-0.8
mm][l]{/}{a}} ({\makebox[-0.8 mm][l]{/}{p}}_1\sigma ^{\mu \nu
}+\sigma ^{\mu \nu } {\makebox[-0.8 mm][l]{/}{p}}_2 )
{\makebox[-0.8 mm][l]{/}{b}}\gamma_{\alpha}
 -\frac{1}{2}\int \overline{d k}
\gamma^{\alpha} {\makebox[-0.8 mm][l]{/}{a}} [({\makebox[-0.8
mm][l]{/}{p}}_1+ {\makebox[-0.8 mm][l]{/}{p}}_2)\sigma ^{\mu \nu
}+\sigma ^{\mu \nu } ({\makebox[-0.8 mm][l]{/}{p}}_1 +
{\makebox[-0.8 mm][l]{/}{p}}_2 )] {\makebox[-0.8
mm][l]{/}{b}}\gamma_{\alpha}.
\end{eqnarray}
 To one-loop order, the inverse of the fermion propagator
reads
\begin{equation}
 S^{-1}_{F}(p_i)={\makebox[-0.8 mm][l]{/}{p}}_i - \Sigma_{(2)}(p_i), \
 \  \ i=1,2,
\end{equation}
where $\Sigma_{(2)}(p_i)$ are one loop self-energy:
\begin{equation}
 \Sigma_{(2)}(p_1) = \int \overline{dk}
\gamma^{\alpha} {\makebox[-0.8 mm][l]{/}{a}} b^2 \gamma_{\alpha},\
\ \  \Sigma_{(2)}(p_2) = \int \overline{dk} \gamma^{\alpha}a^2
{\makebox[-0.8 mm][l]{/}{b}}\gamma_{\alpha}.
\end{equation}
Using Eq.(14) and with some $\gamma$- algebraic calculations,
Eq.(12) leads to

\begin{eqnarray}
& &iq^{\mu} \Lambda^{\nu}_{(2)}(p_1,p_2)
- iq^{\nu} \Lambda^{\mu}_{(2)}(p_1,p_2) \nonumber \\
&=&-\Sigma_{(2)}(p_1)\sigma ^{\mu \nu }- \sigma ^{\mu \nu
}\Sigma_{(2)}(p_2) + (p_{1\lambda }+p_{2\lambda })\varepsilon
^{\lambda \mu \nu \rho }\Lambda _{A\rho(2)}(p_1,p_2) \nonumber \\
& & + \int \overline{d k} \gamma^{\alpha} {\makebox[-0.8
mm][l]{/}{a}} ({\makebox[-0.8 mm][l]{/}{k}}\sigma ^{\mu \nu
}+\sigma ^{\mu \nu } {\makebox[-0.8 mm][l]{/}{k}} ) {\makebox[-0.8
mm][l]{/}{b}}\gamma_{\alpha} \nonumber \\
& & + 2 \int \overline{d k} \{\gamma^{\alpha} {\makebox[-0.8
mm][l]{/}{a}} ({\makebox[-0.8 mm][l]{/}{a}}\sigma ^{\mu \nu
}+\sigma ^{\mu \nu } {\makebox[-0.8 mm][l]{/}{b}} ) {\makebox[-0.8
mm][l]{/}{b}}\gamma_{\alpha}  - (a^2b_{\lambda} +
b^2a_{\lambda})(\gamma ^{\lambda}\sigma ^{\mu \nu} + \sigma ^{\mu
\nu}\gamma ^{\lambda}) \},
\end{eqnarray}
where $\Lambda_{A\rho(2)}(p_1,p_2)= \int \bar{dk}\gamma ^{\alpha}
{\makebox[-0.8 mm][l]{/}{a}}\gamma_{\rho}\gamma_5 {\makebox[-0.8
mm][l]{/}{b}}\gamma_{\alpha}]$. Note that Eq.(15) is same as
Eq.(23) of Ref.[13].

Using the identity $\gamma ^{\lambda}\sigma ^{\mu \nu} = \sigma
^{\mu \nu}\gamma ^{\lambda}+
2i(g^{\lambda\mu}\gamma^{\nu}-g^{\lambda\nu}\gamma^{\mu}) $, and
performing some $\gamma$- algebraic calculations, I obtain from
Eq.(15):
\begin{eqnarray}
& &iq^{\mu} \Lambda^{\nu}_{(2)}(p_1,p_2)
- iq^{\nu} \Lambda^{\mu}_{(2)}(p_1,p_2) \nonumber \\
&=&-\Sigma_{(2)}(p_1)\sigma ^{\mu \nu }- \sigma ^{\mu \nu
}\Sigma_{(2)}(p_2) + (p_{1\lambda }+p_{2\lambda })\varepsilon
^{\lambda \mu \nu \rho }\Lambda _{A\rho(2)}(p_1,p_2) \nonumber \\
& & + \int \overline{d k} \gamma^{\alpha} {\makebox[-0.8
mm][l]{/}{a}} ({\makebox[-0.8 mm][l]{/}{k}}\sigma ^{\mu \nu
}+\sigma ^{\mu \nu } {\makebox[-0.8 mm][l]{/}{k}} ) {\makebox[-0.8
mm][l]{/}{b}}\gamma_{\alpha} \nonumber \\
& & + \int \overline{d k} \{ b^2 \gamma_{\alpha} {\makebox[-0.8
mm][l]{/}{a}} ( \gamma ^{\alpha}\sigma ^{\mu \nu }+\sigma ^{\mu
\nu }\gamma ^{\alpha} )  + a^2 (\gamma^{\alpha}\sigma ^{\mu \nu} +
\sigma ^{\mu \nu}\gamma ^{\alpha}) {\makebox[-0.8
mm][l]{/}{b}}\gamma_{\alpha} \},
\end{eqnarray}
which shows that there is no additional piece $\sim (d-4)$. To
confirm this result, let me make further check by an explicit
calculation in d-dimensions. The possible additional piece
$(d-4)N^{\mu\nu}$, if exits, should be given by the difference
between Eq.(15) and Eq.(16):
\begin{eqnarray}
(d-4)N^{\mu\nu}&=& 2 \int \overline{d k} \{\gamma^{\alpha}
{\makebox[-0.8 mm][l]{/}{a}} ({\makebox[-0.8 mm][l]{/}{a}}\sigma
^{\mu \nu }+\sigma ^{\mu \nu } {\makebox[-0.8 mm][l]{/}{b}} )
{\makebox[-0.8 mm][l]{/}{b}}\gamma_{\alpha}  - (a^2b_{\lambda} +
b^2a_{\lambda})(\gamma ^{\lambda}\sigma
^{\mu \nu} + \sigma ^{\mu \nu}\gamma ^{\lambda}) \}\nonumber \\
& & - \int\overline{d k}\{ b^2 \gamma_{\alpha} {\makebox[-0.8
mm][l]{/}{a}} ( \gamma ^{\alpha}\sigma ^{\mu \nu }+\sigma ^{\mu
\nu }\gamma ^{\alpha} )   + a^2 (\gamma^{\alpha}\sigma ^{\mu \nu}
+ \sigma ^{\mu \nu}\gamma ^{\alpha}) {\makebox[-0.8
mm][l]{/}{b}}\gamma_{\alpha} \}.
\end{eqnarray}
Using the Dirac algebra in d-dimensions
\begin{equation}
\gamma^{\alpha}\gamma^{\mu}\gamma^{\lambda}\gamma^{\nu}\gamma_{\alpha}
= -2\gamma^{\nu}\gamma^{\lambda}\gamma^{\mu}-
(d-4)\gamma^{\mu}\gamma^{\lambda}\gamma^{\nu},  \ \ \
\gamma^{\alpha}\gamma^{\mu}\gamma_{\alpha}= - (d-2)\gamma^{\mu},
\end{equation}
and performing the $\gamma$-algebraic calculations, I find that
$(d-4)N^{\mu\nu} = 0 $, which indicates again that there is no
additional piece $\sim (d-4)$ in Eq.(16).

Combining Eqs.(9)-(13) with (16) gives the transverse WT relation
for the fermion-boson vertex to one-loop order in d-dimensions (
for massless case):
\begin{eqnarray}
& &iq^\mu \Gamma _V^\nu (p_1,p_2)-iq^\nu \Gamma _V^\mu (p_1,p_2)\nonumber \\
&=&S_F^{-1}(p_1)\sigma ^{\mu \nu }+\sigma ^{\mu \nu }S_F^{-1}(p_2)
+(p_{1\lambda }+p_{2\lambda })\varepsilon ^{\lambda \mu \nu \rho }
\Gamma _{A\rho }(p_1,p_2) \nonumber \\
& &-\int \frac{d^{d}k}{(2\pi)^{d}}2k_{\lambda} \varepsilon
^{\lambda \mu \nu \rho }\Gamma _{A\rho }(p_1,p_2;k)
\end{eqnarray}
with
\begin{eqnarray}
& &\int \frac{d^{d}k}{(2 \pi)^{d}}2k_{\lambda}
\varepsilon ^{\lambda \mu \nu \rho }\Gamma _{A\rho }(p_1,p_2;k)\nonumber \\
&=& g^2\int \frac{d^{d}k}{(2 \pi)^{d}}2k_{\lambda} \varepsilon
^{\lambda \mu \nu \rho }\gamma^{\alpha} \frac{1}{{\makebox[-0.8
mm][l]{/}{p}}_1- {\makebox[-0.8
mm][l]{/}{k}}}\gamma_{\rho}\gamma_5 \frac{1}{{\makebox[-0.8
mm][l]{/}{p}}_2-{\makebox[-0.8 mm][l]{/}{k}}}\gamma^{\beta}
\frac{-i}{k^2}[g_{\alpha\beta}+(\xi-1)\frac{k_{\alpha}k_{\beta}}{k^2}]\nonumber \\
& &+g^2\int \frac{d^{d}k}{(2 \pi)^{d}}2 \varepsilon ^{\alpha \mu
\nu \rho }[\gamma^{\beta} \frac{1}{{\makebox[-0.8
mm][l]{/}{p}}_1-{\makebox[-0.8 mm][l]{/}{k}}}\gamma_{\rho}\gamma_5
+ \gamma_{\rho}\gamma_5 \frac{1}{{\makebox[-0.8
mm][l]{/}{p}}_2-{\makebox[-0.8 mm][l]{/}{k}}}\gamma^{\beta}]
\frac{-i}{k^2}[g_{\alpha\beta}+(\xi-1)\frac{k_{\alpha}k_{\beta}}{k^2}]
,
\end{eqnarray}
where the covariant gauge is used to replace the Feynman gauge.
The result shows that Eq.(19) with Eq.(20) given in d-dimensions
has the same form as Eq.(3) with Eq.(4) given in 4-dimensions and
so there is no need for an additional piece $\sim (d-4)$ to
include for the transverse WT relation to hold in 4-dimensions. I
would like to emphasis that the relation (19) together with (20)
are exact to one-loop order because they have been deduced exactly
without any ambiguity.

\section{ Computing terms of the Transverse Ward-Takahashi Relation to One-Loop Order}

Let me check the transverse WT relation (19) with (20) by an
explicit computation of terms in Eq.(19). For simplicity, I
consider the massless fermion case and the Feynman gauge. The main
task is to compute the integrals given by Eq.(20). These integrals
were computed directly in Ref.[13](I will return to discuss the
problem in the computation of Ref.[13] in next section ). Here I
use another way to compute the integrals, that is, at first, using
the identity of $\gamma$ matrices, Eq.(11), and then do the
integral calculations. Thus I find

\begin{eqnarray}
& &- \int \frac{d^{d}k}{(2 \pi)^{d}}2k_{\lambda}
\varepsilon ^{\lambda \mu \nu \rho }\Gamma _{A\rho }(p_1,p_2;k)\nonumber \\
&=& g^2\int \frac{d^{d}k}{(2 \pi)^{d}}\gamma^{\alpha}
\frac{1}{{\makebox[-0.8 mm][l]{/}{p}}_1-{\makebox[-0.8
mm][l]{/}{k}} } ({\makebox[-0.8 mm][l]{/}{p}}_1\sigma^{\mu \nu}+
\sigma^{\mu \nu}{\makebox[-0.8 mm][l]{/}{p}}_2)
\frac{1}{{\makebox[-0.8 mm][l]{/}{p}}_2-{\makebox[-0.8
mm][l]{/}{k}} }\gamma_{\alpha} \frac{-i}{k^2}
\nonumber \\
& & + g^2\int \frac{d^{d}k}{(2 \pi)^{d}} [\gamma_{\alpha}
\frac{1}{{\makebox[-0.8 mm][l]{/}{p}}_1-{\makebox[-0.8
mm][l]{/}{k}} }\gamma^{\alpha}\sigma^{\mu \nu} + \sigma^{\mu
\nu}\gamma^{\alpha} \frac{1}{{\makebox[-0.8
mm][l]{/}{p}}_2-{\makebox[-0.8 mm][l]{/}{k}} }\gamma_{\alpha}]
\frac{-i}{k^2},
\end{eqnarray}
which may be denoted as $P_4^{\mu\nu}$. The computation gives

\begin{eqnarray}
P_4^{\mu\nu} &=& -\frac{i\alpha}{2 \pi^3}\{ {\makebox[-0.8
mm][l]{/}{p}}_2 ( {\makebox[-0.8 mm][l]{/}{p}}_2\sigma^{\mu
\nu}+\sigma^{\mu \nu} {\makebox[-0.8 mm][l]{/}{p}}_1 )
{\makebox[-0.8 mm][l]{/}{p}}_1 J^{(0)} - [ {\makebox[-0.8
mm][l]{/}{p}}_2 ( {\makebox[-0.8 mm][l]{/}{p}}_2\sigma^{\mu
\nu}+\sigma^{\mu \nu} {\makebox[-0.8 mm][l]{/}{p}}_1 )
\gamma^{\lambda} \nonumber \\
& &+ \gamma^{\lambda} ( {\makebox[-0.8 mm][l]{/}{p}}_2\sigma^{\mu
\nu} +\sigma^{\mu \nu}{\makebox[-0.8 mm][l]{/}{p}}_1 )
{\makebox[-0.8 mm][l]{/}{p}}_1 ] J^{(1)}_{\lambda}
+\gamma^{\lambda}( {\makebox[-0.8 mm][l]{/}{p}}_2\sigma^{\mu \nu}
+\sigma^{\mu \nu}{\makebox[-0.8 mm][l]{/}{p}}_1  )\gamma^{\eta}
J^{(2)}_{\lambda \eta} \nonumber \\
& &+ (d-4)\frac{i\alpha}{4 \pi^3}\{ {\makebox[-0.8 mm][l]{/}{p}}_1
( {\makebox[-0.8 mm][l]{/}{p}}_1\sigma^{\mu \nu}+\sigma^{\mu \nu}
{\makebox[-0.8 mm][l]{/}{p}}_2 ) {\makebox[-0.8 mm][l]{/}{p}}_2
J^{(0)} - [ {\makebox[-0.8 mm][l]{/}{p}}_1 ( {\makebox[-0.8
mm][l]{/}{p}}_1\sigma^{\mu \nu}+\sigma^{\mu \nu} {\makebox[-0.8
mm][l]{/}{p}}_2 )
\gamma^{\lambda} \nonumber \\
& &+ \gamma^{\lambda} ( {\makebox[-0.8 mm][l]{/}{p}}_1\sigma^{\mu
\nu} +\sigma^{\mu \nu}{\makebox[-0.8 mm][l]{/}{p}}_2 )
{\makebox[-0.8 mm][l]{/}{p}}_2 ]J^{(1)}_{\lambda}
+\gamma^{\lambda}( {\makebox[-0.8 mm][l]{/}{p}}_1\sigma^{\mu \nu}
+\sigma^{\mu \nu}{\makebox[-0.8 mm][l]{/}{p}}_2  )\gamma^{\eta}
J^{(2)}_{\lambda \eta} \nonumber \\
& &+\Sigma_{(2)}(p_1)\sigma ^{\mu \nu }+\sigma ^{\mu \nu
}\Sigma_{(2)}(p_2) ,
\end{eqnarray}
where
\begin{equation}
\Sigma_{(2)}(p_i) = \frac{i\alpha}{4 \pi^3}(d-2)[{\makebox[-0.8
mm][l]{/}{p}}_i
K^{(0)}(p_i)-\gamma^{\lambda}K^{(1)}_{\lambda}(p_i)], \ \ \ i=1,2.
\end{equation}
Here $\alpha=g^2/4\pi$, $J^{(0)}$, $J^{(1)}_{\lambda}$,
 $J^{(2)}_{\lambda \eta}$, $K^{(0)}(p_i)$ and $K^{(1)}_{\lambda}(p_i)$  are some integrals
 listed in Appendix.

Now let me compute other terms in the transverse WT relation (19)
by following notations: $P_1^{\mu\nu}=iq^{\mu}
\Gamma^{\nu}_{V(2)}(p_1,p_2) -iq^{\nu}
\Gamma^{\mu}_{V(2)}(p_1,p_2)$,
$P_2^{\mu\nu}=S_{F(2)}^{-1}(p_1)\sigma ^{\mu \nu }+\sigma ^{\mu
\nu }S_{F(2)}^{-1}(p_2)$, $P_3^{\mu\nu}=(p_{1\lambda }+p_{2\lambda
})\varepsilon ^{\lambda \mu \nu \rho } \Gamma _{A\rho(2)
}(p_1,p_2)$.

The fermion-boson vertex function at one-loop order is familiar:
\begin{eqnarray}
 \Lambda^{\mu}_{(2)}(p_1,p_2)
&=& - \frac{i\alpha}{4 \pi^3}\{ \gamma^{\alpha}{\makebox[-0.8
mm][l]{/}{p}}_1\gamma^{\mu}{\makebox[-0.8 mm][l]{/}{p}}_2
\gamma_{\alpha}J^{(0)} \nonumber \\
& &- (\gamma^{\alpha}{\makebox[-0.8
mm][l]{/}{p}}_1\gamma^{\mu}\gamma^{\lambda} \gamma_{\alpha}
+\gamma^{\alpha}\gamma^{\lambda}\gamma^{\mu}{\makebox[-0.8
mm][l]{/}{p}}_2\gamma_{\alpha}) J^{(1)}_{\lambda}
+\gamma^{\alpha}\gamma^{\lambda}\gamma^{\mu}\gamma^{\eta}\gamma_{\alpha}
J^{(2)}_{\lambda \eta}\} .
\end{eqnarray}
Using Eq.(24) and Eq.(18), it is straightforward to get
\begin{eqnarray}
P_1^{\mu\nu} &=& iq^{\mu}\gamma^{\nu}-iq^{\nu}\gamma^{\mu}-
\frac{\alpha}{2 \pi^3}\{ {\makebox[-0.8mm][l]{/}{p}}_2
(q^{\mu}\gamma^{\nu}-q^{\nu}\gamma^{\mu})
{\makebox[-0.8 mm][l]{/}{p}}_1J^{(0)} \nonumber \\
& &- [{\makebox[-0.8mm][l]{/}{p}}_2
(q^{\mu}\gamma^{\nu}-q^{\nu}\gamma^{\mu}) \gamma^{\lambda}
+\gamma^{\lambda}(q^{\mu}\gamma^{\nu}-q^{\nu}\gamma^{\mu})
{\makebox[-0.8 mm][l]{/}{p}}_1]J^{(1)}_{\lambda} +\gamma^{\lambda}
(q^{\mu}\gamma^{\nu}-q^{\nu}\gamma^{\mu})
\gamma^{\eta}J^{(2)}_{\lambda \eta} \}\nonumber \\
& &-(d-4) \frac{\alpha}{4 \pi^3}\{ {\makebox[-0.8mm][l]{/}{p}}_1
(q^{\mu}\gamma^{\nu}-q^{\nu}\gamma^{\mu}) {\makebox[-0.8
mm][l]{/}{p}}_2J^{(0)} - [{\makebox[-0.8mm][l]{/}{p}}_1
(q^{\mu}\gamma^{\nu}-q^{\nu}\gamma^{\mu}) \gamma^{\lambda}
+\gamma^{\lambda}(q^{\mu}\gamma^{\nu}-q^{\nu}\gamma^{\mu})
{\makebox[-0.8 mm][l]{/}{p}}_2]J^{(1)}_{\lambda}
\nonumber \\
& &+\gamma^{\lambda} (q^{\mu}\gamma^{\nu}-q^{\nu}\gamma^{\mu})
\gamma^{\eta}J^{(2)}_{\lambda \eta} \} .
\end{eqnarray}

The axial-vector vertex function at one-loop order can be obtained
from Eq.(24) by replacing $\gamma^{\mu}$ with
$\gamma^{\mu}\gamma_5$, thus it gives
\begin{eqnarray}
P_3^{\mu\nu} &=& (p_{1\lambda }+p_{2\lambda })\varepsilon
^{\lambda \mu \nu \rho } \gamma _{\rho}\gamma_5  \nonumber \\
 & &+\frac{i\alpha}{4\pi^3}\{ {\makebox[-0.8 mm][l]{/}{p}}_2\{
 {\makebox[-0.8 mm][l]{/}{p}}_1 +{\makebox[-0.8 mm][l]{/}{p}}_2,
\sigma^{\mu \nu} \} {\makebox[-0.8 mm][l]{/}{p}}_1 J^{(0)} - [
{\makebox[-0.8 mm][l]{/}{p}}_2 \{ {\makebox[-0.8 mm][l]{/}{p}}_1
+{\makebox[-0.8 mm][l]{/}{p}}_2, \sigma^{\mu \nu} \}
\gamma^{\lambda} + \gamma^{\lambda} \{ {\makebox[-0.8
mm][l]{/}{p}}_1 +{\makebox[-0.8 mm][l]{/}{p}}_2, \sigma^{\mu \nu}
 \} {\makebox[-0.8 mm][l]{/}{p}}_1 ] J^{(1)}_{\lambda} \nonumber \\
 & & +\gamma^{\lambda}\{{\makebox[-0.8 mm][l]{/}{p}}_1 +
{\makebox[-0.8 mm][l]{/}{p}}_2, \sigma^{\mu \nu} \}\gamma^{\eta}
J^{(2)}_{\lambda \eta} \}\nonumber \\
& & -(d-4)\frac{i\alpha}{8\pi^3}\{ {\makebox[-0.8
mm][l]{/}{p}}_1\{
 {\makebox[-0.8 mm][l]{/}{p}}_1 +{\makebox[-0.8 mm][l]{/}{p}}_2,
\sigma^{\mu \nu} \} {\makebox[-0.8 mm][l]{/}{p}}_2 J^{(0)} - [
{\makebox[-0.8 mm][l]{/}{p}}_1 \{ {\makebox[-0.8 mm][l]{/}{p}}_1
+{\makebox[-0.8 mm][l]{/}{p}}_2, \sigma^{\mu \nu} \}
\gamma^{\lambda} + \gamma^{\lambda} \{ {\makebox[-0.8
mm][l]{/}{p}}_1 +{\makebox[-0.8 mm][l]{/}{p}}_2, \sigma^{\mu \nu}
 \} {\makebox[-0.8 mm][l]{/}{p}}_2 ] J^{(1)}_{\lambda} \nonumber \\
 & & +\gamma^{\lambda}\{{\makebox[-0.8 mm][l]{/}{p}}_1 +
{\makebox[-0.8 mm][l]{/}{p}}_2, \sigma^{\mu \nu} \}\gamma^{\eta}
J^{(2)}_{\lambda \eta}\}.
\end{eqnarray}
At last, it is easy to get
\begin{equation}
P_2^{\mu\nu}= {\makebox[-0.8 mm][l]{/}{p}}_1\sigma^{\mu
\nu}+\sigma^{\mu \nu} {\makebox[-0.8 mm][l]{/}{p}}_2
-\Sigma_{(2)}(p_1)\sigma ^{\mu \nu }- \sigma ^{\mu \nu
}\Sigma_{(2)}(p_2).
\end{equation}

Now using the identity (11) and Eq.(9), I obtain
\begin{equation}
P_1^{\mu\nu}- P_2^{\mu\nu}- P_3^{\mu\nu}- P_4^{\mu\nu} = 0.
\end{equation}
This shows that the transverse WT relation for the fermion-boson
vertex to one-loop order is satisfied indeed and there is no need
for an additional piece proportional to $(d-4)$ to include, which
confirms the conclusion obtained in last section.

\section{How a so-called additional piece might be separated out ? }

In a recent paper, Pennington and Willams[13] claimed that an
additional piece must be included in the evaluation of the Wilson
line component(i.e. the integral-term) for the transverse WT
relation to hold in 4-dimensions as(see Eq.(25) of Ref.[13]):
\begin{eqnarray}
& &iq^\mu \Gamma _V^\nu (p_1,p_2)-iq^\nu \Gamma _V^\mu (p_1,p_2)\nonumber \\
&=&S_F^{-1}(p_1)\sigma ^{\mu \nu }+\sigma ^{\mu \nu }S_F^{-1}(p_2)
+(p_{1\lambda }+p_{2\lambda })\varepsilon ^{\lambda \mu \nu \rho }
\Gamma _{A\rho }(p_1,p_2) \nonumber \\
& &-\int \frac{d^{d}k}{(2\pi)^{d}}2k_{\lambda} \varepsilon
^{\lambda \mu \nu \rho }\tilde{\Gamma }_{A\rho }(p_1,p_2;k)_{P-M} \nonumber \\
& & -4(d-4)g^2\int \frac{d^{d}k}{(2\pi)^{d}}\frac {-i}{k^2}\frac
{(p_1 -k)_{\lambda} \varepsilon ^{\lambda \mu \nu \rho }\gamma
_{\rho }\gamma_5}{(p_1 - k)^2},
\end{eqnarray}
where the last term is the so-called additional piece denoted as
$(d-4)N^{\mu\nu}$ in Ref.[13]. To analyze how they write this form
in Ref.[13], let me follow their relative derivation. The key step
is the derivation from Eq.(23) to Eq.(24) in Ref.[13]. Eq.(23) of
Ref.[13] is same as Eq.(15) of present work. Rearranging Eq.(15)
can gave

\begin{eqnarray}
& &iq^{\mu} \Lambda^{\nu}_{(2)}(p_1,p_2)
- iq^{\nu} \Lambda^{\mu}_{(2)}(p_1,p_2) \nonumber \\
&=&-\Sigma_{(2)}(p_1)\sigma ^{\mu \nu }- \sigma ^{\mu \nu
}\Sigma_{(2)}(p_2) + (p_{1\lambda }+p_{2\lambda })\varepsilon
^{\lambda \mu \nu \rho }\Lambda _{A\rho(2)}(p_1,p_2) \nonumber \\
& & + \int \overline{d k}\gamma^{\alpha} {\makebox[-0.8
mm][l]{/}{a}} ({\makebox[-0.8 mm][l]{/}{k}}\sigma ^{\mu \nu
}+\sigma ^{\mu \nu } {\makebox[-0.8 mm][l]{/}{k}} ) {\makebox[-0.8
mm][l]{/}{b}}\gamma_{\alpha} \nonumber \\
& & - \int \overline{d k}(\gamma^{\alpha}\sigma ^{\mu \nu }+\sigma
^{\mu \nu }\gamma^{\alpha})( a^2 {\makebox[-0.8 mm][l]{/}{b}}-
{\makebox[-0.8 mm][l]{/}{a}}b^2 )\gamma_{\alpha} \nonumber \\
& & - 2(d-4)\int \overline{d k}( {\makebox[-0.8
mm][l]{/}{a}}\sigma ^{\mu \nu} + \sigma ^{\mu \nu}{\makebox[-0.8
mm][l]{/}{a}})b^2.
\end{eqnarray}
This is the same equation as Eq.(24) of Ref.[13]. Combining
Eq.(30) with Eqs.(9) and (10), then leads to the expression (29),
where a modifying integral-term is defined as
\begin{eqnarray}
& &\int \frac{d^{d}k}{(2 \pi)^{d}}2k_{\lambda}
\varepsilon ^{\lambda \mu \nu \rho }\tilde{\Gamma} _{A\rho }(p_1,p_2;k)_{P-M}\nonumber \\
&=& - \int \overline{d k} \gamma^{\alpha} {\makebox[-0.8
mm][l]{/}{a}} ({\makebox[-0.8 mm][l]{/}{k}}\sigma ^{\mu \nu
}+\sigma ^{\mu \nu } {\makebox[-0.8 mm][l]{/}{k}} ) {\makebox[-0.8
mm][l]{/}{b}}\gamma_{\alpha} \nonumber \\
& & + \int \overline{d k}(\gamma^{\alpha}\sigma ^{\mu \nu }+\sigma
^{\mu \nu }\gamma^{\alpha})( a^2 {\makebox[-0.8 mm][l]{/}{b}}-
{\makebox[-0.8 mm][l]{/}{a}}b^2 )\gamma_{\alpha} \nonumber \\
 &=& g^2\int \frac{d^{d}k}{(2 \pi)^{d}}2k_{\lambda} \varepsilon
^{\lambda \mu \nu \rho }\gamma^{\alpha} \frac{1}{{\makebox[-0.8
mm][l]{/}{p}}_1- {\makebox[-0.8
mm][l]{/}{k}}}\gamma_{\rho}\gamma_5 \frac{1}{{\makebox[-0.8
mm][l]{/}{p}}_2-{\makebox[-0.8 mm][l]{/}{k}}}\gamma_{\alpha}
\frac{-i}{k^2}\nonumber \\
& &+g^2\int \frac{d^{d}k}{(2 \pi)^{d}}2 \varepsilon ^{\alpha \mu
\nu \rho }\gamma_{\rho}\gamma_5[ \frac{1}{{\makebox[-0.8
mm][l]{/}{p}}_1-{\makebox[-0.8 mm][l]{/}{k}}}-
\frac{1}{{\makebox[-0.8 mm][l]{/}{p}}_2-{\makebox[-0.8
mm][l]{/}{k}}}]\gamma^{\alpha} \frac{-i}{k^2} ,
\end{eqnarray}
which is obviously different from the integral-term given by
Eq.(20) if comparing the second part of the right-hand side of
Eq.(31) with that of Eq.(20) (in the Feynman gauge). Furthermore,
comparing Eqs.(16) and (19) with Eqs.(30) and (29), I then obtain
following relation:

\begin{eqnarray}
& &\int \frac{d^{d}k}{(2 \pi)^{d}}2k_{\lambda}
\varepsilon ^{\lambda \mu \nu \rho }\Gamma _{A\rho }(p_1,p_2;k)\nonumber \\
&=& \int \frac{d^{d}k}{(2 \pi)^{d}}2k_{\lambda}
\varepsilon ^{\lambda \mu \nu \rho }\tilde{\Gamma} _{A\rho }(p_1,p_2;k)_{P-M}\nonumber \\
& & +4(d-4)g^2\int \frac{d^{d}k}{(2\pi)^{d}}\frac {-i}{k^2}\frac
{(p_1 -k)_{\lambda} \varepsilon ^{\lambda \mu \nu \rho }\gamma
_{\rho }\gamma_5}{(p_1 - k)^2}.
\end{eqnarray}

Thus, it shows clearly that the authors of Ref.[13] separated out
a so-called additional piece $(d-4)N^{\mu\nu}$ from the
integral-term by defining a modifying integral-term, as shown by
Eqs.(31)-(32), which, of cause, does not change the formula of the
transverse WT relation to one-loop order as given by Eq.(19) with
Eq.(20).

However, in the explicit one-loop computation of terms in the
transverse WT relation given in Sec.3 of Ref.[13], the authors of
Ref.[13] replaced the modified integral-term with the original
integral-term ( see Eq.(42) of Ref.[13]) but still included the
additional piece $(d-4)N^{\mu\nu}$. Such a computation is
obviously inconsistent. In fact, there was a problem in their
computation of the second part of the integral-term given in
Ref.[13]. To see what is the problem, let me compute this part in
the Feynman gauge in the following:
\begin{eqnarray}
& &- \int \frac{d^{d}k}{(2 \pi)^{d}}2k_{\lambda}
\varepsilon ^{\lambda \mu \nu \rho }\Gamma _{A\rho }(p_1,p_2;k)[second- part]\nonumber \\
&=& - g^2\int \frac{d^{d}k}{(2 \pi)^{d}}2 \varepsilon ^{\alpha \mu
\nu \rho }[\gamma^{\alpha} \frac{1}{{\makebox[-0.8
mm][l]{/}{p}}_1-{\makebox[-0.8 mm][l]{/}{k}}}\gamma_{\rho}\gamma_5
+ \gamma_{\rho}\gamma_5 \frac{1}{{\makebox[-0.8
mm][l]{/}{p}}_2-{\makebox[-0.8 mm][l]{/}{k}}}\gamma^{\alpha}]
\frac{-i}{k^2} \nonumber \\
& = &\frac {\alpha}{4\pi^3}[\gamma_{\alpha} \gamma^{\lambda}
\gamma^{\alpha\mu\nu}\int d^{d}k \frac {(p_1-k)_{\lambda}}{(p_1 -
k)^2k^2} +\gamma^{\alpha\mu\nu}\gamma^{\lambda}\gamma_{\alpha}
\int d^{d}k \frac {(p_2-k)_{\lambda}}{(p_2 - k)^2k^2} ],
\end{eqnarray}
where the notation $\gamma^{\alpha\mu\nu}$ is given in Eq.(11).
Noticing that
\begin{equation}
\gamma_{\alpha} \gamma^{\lambda} \gamma^{\alpha\mu\nu} = -
\gamma^{\alpha\mu\nu}\gamma^{\lambda}\gamma_{\alpha} + 2i
(d-4)(\gamma ^{\lambda}\sigma ^{\mu \nu} + \sigma ^{\mu \nu}\gamma
^{\lambda}),
\end{equation}
and using this relation into Eq.(33), I find
\begin{eqnarray}
& &- \int \frac{d^{d}k}{(2 \pi)^{d}}2k_{\lambda}
\varepsilon ^{\lambda \mu \nu \rho }\Gamma _{A\rho }(p_1,p_2;k)[second- part]\nonumber \\
&=& -\frac {\alpha}{2\pi^3}[p_{2\lambda}K^{(0)}(0,p_2) -
p_{1\lambda}K^{(0)}(p_1,0)-K^{(1)}_{\lambda}(0,p_2)+K^{(1)}_{\lambda}(p_1,0)]\nonumber \\
& &\times
(2\gamma^{\mu}g^{\lambda\nu}-2\gamma^{\nu}g^{\lambda\mu}+
(d-4)(\gamma^{\mu}\gamma^{\nu}- g^{\mu\nu})\gamma^{\lambda}) \nonumber \\
& &+\frac
{i\alpha}{2\pi^3}(d-4)[p_{1\lambda}K^{(0)}(p_1,0)-K^{(1)}_{\lambda}(p_1,0)]
(\gamma ^{\lambda}\sigma ^{\mu \nu} +
\sigma^{\mu\nu}\gamma^{\lambda}),
\end{eqnarray}
where the first piece of the right-hand side of Eq.(35) gives the
corresponding result of Ref.[13], while the contribution of second
piece $\sim (d-4)$ was missing in the corresponding computation of
Ref.[13](see Eq.(44) of Ref.[13]). This missing piece by Ref.[13]
in the integral form is
\begin{equation}
-4(d-4)g^2\int \frac{d^{d}k}{(2\pi)^{d}}\frac {-i}{k^2}\frac {(p_1
-k)_{\lambda} \varepsilon ^{\lambda \mu \nu \rho }\gamma _{\rho
}\gamma_5}{(p_1 - k)^2},
\end{equation}
which is just equal to the so-called additional piece given in
Eq.(29). Thus this computation indicates clearly that a piece
equal to the so-called additional piece was missing in the
computation of the integral-term in Ref.[13]. Including such a
missing piece into their computation of the integral-term, then an
additional piece does not need to be introduced.

\section{Conclusion and Remark}

In conclusion, I present the complete expression of the
integral-term involving the non-local axial-vector vertex function
and hence give the general expression of the transverse
Ward-Takahashi (WT) relation for the fermion-boson vertex function
in momentum space in 4-dimensional QED, which should be satisfied
both perturbatively and non-perturbatively like the normal WT
identity. I have deduced that this transverse WT relation to
one-loop order in d-dimensions, with $d=4+\epsilon$, has the same
form as one given in 4-dimensions and there is no need for an
additional piece proportional to $(d-4)$ to include in the
evaluation of the integral-term for this relation to hold in
4-dimensions. This result has been confirmed by an explicit
computation of terms in this transverse WT relation to one-loop
order in the Feynman gauge and the massless fermion case.

It has been shown that the authors of Ref.[13] in their paper
separated out a so-called additional piece, $(d-4)N^{\mu\nu}$,
from the integral-term involving the non-local axial-vector vertex
by defining a modified integral-term, which in fact does not
change the original formula of the transverse WT relation to one
loop order. Thus it needs not have introduced such an additional
piece.

The transverse WT relation for the fermion-boson vertex, Eq.(3),
shows that the transverse part of the fermion-boson (vector)
vertex is related to the axial-vector and tensor vertices. Thus in
order to constrain completely the fermion-boson vertex, it is
needed to build the transverse WT relations for the axial-vector
and tensor vertices as well. Their expressions in coordinate space
have been derived in Ref.[9], where, however, the corresponding
expressions in momentum space neglected the contributions from
integral-terms. The complete expressions of the transverse WT
relations for the axial-vector and tensor vertices in momentum
space and their expressions to one-loop order can be obtained by a
similar procedure given in this paper. Then the full fermion-boson
vertex can be derived consistently in terms of a set of the normal
and transverse WT relations for the fermion vertex functions,
which will be discussed elsewhere [14].

\section*{Appendix}

\begin{equation}
 J^{(0)}(p_1,p_2) = \int_M d^{4}k\frac{1}{k^2( p_1-k )^2( p_2-k )^2} ,
\end{equation}
\begin{equation}
 J^{(1)}_{\mu}(p_1,p_2) = \int_M d^{4}k\frac{k_{\mu}}{k^2( p_1-k )^2(p_2-k )^2} ,
\end{equation}
\begin{equation}
 J^{(2)}_{\mu \nu}(p_1,p_2) = \int_M d^{4}k\frac{k_{\mu}k_{\nu}}
{k^2( p_1-k )^2 ( p_2-k )^2 } ,
\end{equation}
\begin{equation}
 K^{(0)}(p_i) = \int_M d^{4}k\frac{1}{( p_i-k )^2 k^2 } , \ \
 \i=1,2,
\end{equation}
\begin{equation}
 K^{(1)}_{\mu}(p_i) = \int_M d^{4}k\frac{k_{\mu}}{( p_i-k )^2 k^2 }
 , \ \ \ i=1,2,
\end{equation}
where $K^{(0)}(p_1)=K^{(0)}(p_1,0)$,
$K^{(0)}(p_2)=K^{(0)}(0,p_2)$, $K^{(1)}_{\mu}(p_1)=
K^{(1)}_{\mu}(p_1,0)$, $ K^{(1)}_{\mu}(p_2)=
K^{(1)}_{\mu}(0,p_2)$. These integrals can be carried out in the
cutoff regularization scheme or in the dimensional regularization
scheme[4,5,6,13] . Here I do not list the results of these
integrals since I do not need to use these detailed results in
this work.

\section*{Acknowledgments}

I would like to thank H.S.Zong for telling me Ref.[13] and useful
conversations during a recent CCAST workshop (March 6-10,2006,
Beijing, China). This work is supported by the National Natural
Science Foundation of China under grant No.90303006.

\end{document}